\documentclass[pra,reprint,nofootinbib,showpacs,superscriptaddress,twocolumn]{revtex4-1}

\usepackage{graphicx}
\usepackage{hyperref}
\usepackage{amsfonts}
\usepackage{amsmath,amssymb}
\usepackage{bm}
\usepackage{color}
\usepackage{longtable}

\usepackage{algorithmicx}
\usepackage[ruled,vlined]{algorithm2e}
\SetAlgoCaptionSeparator{ }

\def\be{\begin{equation}}
\def\ee{\end{equation}}
\def\bea{\begin{eqnarray}}
\def\eea{\end{eqnarray}}
\def\ba{\begin{array}}
\def\ea{\end{array}}
\def\bem{\begin{multline}}
\def\eem{\end{multline}}

\newcommand{\fref}[1]{Fig.~\ref{#1}}

\newcommand{\tref}[1]{Table~\ref{#1}}
\newcommand{\sref}[1]{Section~\ref{#1}}

\providecommand{\U}[1]{\protect\rule{.1in}{.1in}}

\begin{document}
\title{Decentralized Base-Graph Routing for the Quantum Internet}

\author{Laszlo Gyongyosi}
\affiliation{School of Electronics and Computer Science, University of Southampton, Southampton, SO17 1BJ, UK}
\email{l.gyongyosi@soton.ac.uk}
\affiliation{Department of Networked Systems and Services, Budapest University of Technology and Economics, Budapest, H-1117 Hungary}
\affiliation{MTA-BME Information Systems Research Group, Hungarian Academy of Sciences, Budapest, H-1051 Hungary}
\author{Sandor Imre}
\affiliation{Department of Networked Systems and Services, Budapest University of Technology and Economics, Budapest, H-1117 Hungary}

\begin{abstract}
Quantum repeater networks are a fundamental of any future quantum Internet and long-distance quantum communications. The entangled quantum nodes can communicate through several different levels of entanglement, leading to a heterogeneous, multi-level network structure. The level of entanglement between the quantum nodes determines the hop distance and the probability of the existence of an entangled link in the network. Here, we define a decentralized routing for entangled quantum networks. The proposed method allows an efficient routing to find the shortest paths in entangled quantum networks by using only local knowledge of the quantum nodes. We give bounds on the maximum value of the total number of entangled links of a path. The proposed scheme can be directly applied in practical quantum communications and quantum networking scenarios.   
\end{abstract}

\maketitle

\section{Introduction}
In the quantum Internet \cite{ref1, ref3, ref4, ref7}, the quantum nodes are connected with each other through entangled links \cite{ref1,ref2, ref3, ref4, ref5, ref6, ref7, ref8} allowing one to perform quantum communications beyond the fundamental limits of traditional sender-receiver communications \cite{ref48,ref49,ref50}. The entangled quantum nodes can share several different levels of entanglement, leading to a heterogeneous, \textit{multi-level entanglement} network structure \cite{ref1,ref8, ref15, ref16, ref17, ref18, ref19, ref20, ref21, ref22, ref23, ref24, ref25, ref26, ref27, ref28}. The \textit{level of entanglement} between the quantum nodes determines the achievable hop distance, the number of spanned intermediate nodes, and the probability of the existence of an entangled link \cite{ref29,ref30,ref31,ref32,ref33,ref34,ref35,ref36,ref37,ref38,ref39,ref40, ref47, ref50,ref51,ref52,ref53,ref54,ref55}. For an ${{\text{L}}_{l}}$-level entangled link, the hop distance between quantum nodes $x$ and $y$ is ${{2}^{l-1}}$, and each ${{\text{L}}_{l}}$-level entangled link $E\left( x,y \right)$ can be established only with a given probability, $0<{{\Pr }_{{{\text{L}}_{l}}}}\left( E\left( x,y \right) \right)\le 1$, which depends on the properties of the actual overlay quantum network \cite{ref1,ref2, ref3, ref4, ref5, ref6, ref7, ref8, ref28, ref29, ref30, ref31, ref32, ref33, ref34}. As the level of entanglement increases, the number of spanned nodes also increases, which decreases the probability of the existence of a higher-level entangled link in the network  \cite{ref1,ref8, ref15, ref25, ref26, ref27, ref28, ref29, ref30, ref31, ref32, ref33, ref34}. Note that each quantum node can have an arbitrary number of entangled node contacts with an arbitrary level of entanglement between them. The intermediate nodes between $x$ and $y$ are referred to as quantum repeater nodes and participate only in the process of entanglement distribution from $x$ to $y$.

In an entangled quantum network with heterogeneous entanglement levels, finding the shortest path between arbitrary quantum nodes for the level of entanglement is a crucial task to transmit a message between the nodes in as few steps as possible. Since in practical scenarios there is no global knowledge available about the nodes or about the properties of the entangled links, the routing has to be performed in a \textit{decentralized} manner. In particular, our decentralized routing uses only local knowledge about the nodes and their neighbors and their shared level of entanglement.

Here we show that the probability that a specific level of entanglement exists between the quantum nodes in the entangled overlay quantum network $N$ is proportional to the L1 distance of the nodes in an $n$-sized base-graph. While most of the currently available quantum routing methods \cite{ref1,ref8, ref28, ref29, ref30, ref31, ref32, ref33, ref34} represent a variant of Dijkstra's shortest path algorithm \cite{ref41}, the efficiency of these routing approaches is limited. We have found that the probability distribution of the entangled links can be described by an inverse $k$-power distribution, where $k$ is the dimension of the base-graph $G^k$, making it possible to achieve an $\mathcal{O}{{\left( \log n \right)}^{2}}$ decentralized routing in an entangled overlay quantum network. A $k$-dimensional base-graph contains all quantum nodes and entangled links of the overlay quantum network via a set of nodes and edges such that each link preserves the level of entanglement and corresponding probabilities. Specifically, the construction of the base-graph of an entangled overlay network is a challenge, since in a practical decentralized networking scenario, there is no global knowledge about the exact local positions of the nodes or other coordinates. Particularly, mapping from the entangled overlay quantum network to a base-graph has to be achieved without revealing any routing-related information by security assumptions.  It is necessary to embed the entangled overlay quantum network with the probabilistic entangled links onto a simple base-graph if we want to achieve an efficient decentralized routing. Note, that the quantum links are assumed to be probabilistic, since in a quantum repeater network, both the entanglement purification and the entanglement swapping procedures are probabilistic processes \cite{ref1,ref2, ref3, ref4, ref5, ref6, ref7, ref8}. As follows, quantum entanglement between the distant points can exist only with a given probability, and this probability further decreased by the noise of the physical links used for the transmission.

As we show by utilizing sophisticated mathematical tools, the problem of embedding can be reduced to a statistical estimation task, and thus the base-graph can be prepared for the decentralized routing. Therefore, the shortest path in the heterogeneous entanglement levels of the quantum network can be determined by the L1 metric in the base-graph. Precisely, since the probability of a high-level entangled link between the nodes is lower than the probability of a low-level entanglement, we can assign positions to the quantum nodes in the base-graph according to the \textit{a posteriori} distribution of the positions.
 
The system model allows the utilization of both bipartite and multipartite entangled states. It is because, while for a bipartite entangled system the entangled link is directly formulated between the two quantum systems, in the case of a multipartite entangled system the entangled links are formulated between the entangled partitions of the multipartite entangled state in the network model. 

We show that the proposed method can be applied for an arbitrary-sized entangled quantum network, and by utilizing entangled links, our decentralized routing does not require transmission of any routing-related information in the network. We also reveal the diameter bounds of a multi-level entangled quantum network, where the diameter refers to the maximum value of the shortest path (the total number of entangled links in a path) between a source and a target quantum node.

The contributions of our manuscript are as follows: 
\begin{enumerate}
\item  \textit{We define a decentralized routing for the quantum Internet. We construct a special graph, called base-graph, that contains all information about the quantum network to perform a high performance routing.}
\item \textit{We show that the probability distribution of the entangled links can be modeled by a specific distribution in a base-graph.}
\item \textit{The proposed method allows us to perform efficient routing to find the shortest paths in entangled quantum networks by using only local knowledge of the quantum nodes.}
\item \textit{We derive the computational complexity of the proposed routing scheme.}
\item \textit{We give bounds on the maximum value of the total number of entangled links of the path.}
\end{enumerate}

This paper is organized as follows. In \sref{sec2}, the proposed decentralized routing approach is discussed. \sref{sec3} provides the computational complexity of the scheme. In \sref{sec4} the diameter bounds are derived. Finally, \sref{con} concludes the paper.

\section{System Model}
\label{sec2}
Let us formalize our statements in a strict mathematical manner. Let $V$ refer to the nodes of an overlay entangled quantum network $N$, which consists of a transmitter node $A\in V$, a receiver node $B\in V$, and quantum repeater nodes $R_i\in V$, $i=1,\dots ,q$. Let $E=\left\{E_j\right\}$, $j=1,\dots ,m$ refer to a set of edges between the nodes of $V$, where each $E_j$ identifies an $\text{L}_l$-level entanglement, $l=1,\dots ,r$, between quantum nodes $x_j$ and $y_j$ of edge $E_j$, respectively.

An $N=\left(V,E\right)$ overlay quantum repeater network consists of several single-hop and multi-hop entangled nodes, such that the single-hop entangled nodes are directly connected through an $\text{L}_1$-level entanglement, while the multi-hop entangled nodes communicate through $\text{L}_l$-level entanglement. According to the working mechanism of a doubling quantum repeater architecture \cite{ref1,ref2,ref3,ref4}, the number of spanned nodes is doubled in each level ${{l}_{sw}}=l-1$ of entanglement swapping. Therefore, the $d{\left(x,y\right)}_{\text{L}_l}$ hop distance in $N$ for the $\text{L}_l$-level entangled nodes $x,y\in V$ is denoted by 
\begin{equation} \label{eq1} 
d{\left(x,y\right)}_{\text{L}_l}=2^{l-1}, 
\end{equation} 
 with $d{\left(x,y\right)}_{\text{L}_l}-1$ intermediate nodes between the nodes $x$ and $y$. Thus, $l=1$ refers to a direct quantum link connection between two quantum nodes $x$ and $y$ without intermediate quantum repeaters. The probability that an $\text{L}_l$-level entangled link $E\left(x,y\right)$ exists between $x,y\in V$ is ${\Pr }_{\text{L}_l}\left(E\left(x,y\right)\right)$, which depends on the actual network.

\begin{center}
\begin{figure*}[!htbp]
\begin{center}
 	 \includegraphics[angle = 0,width=0.7\linewidth]{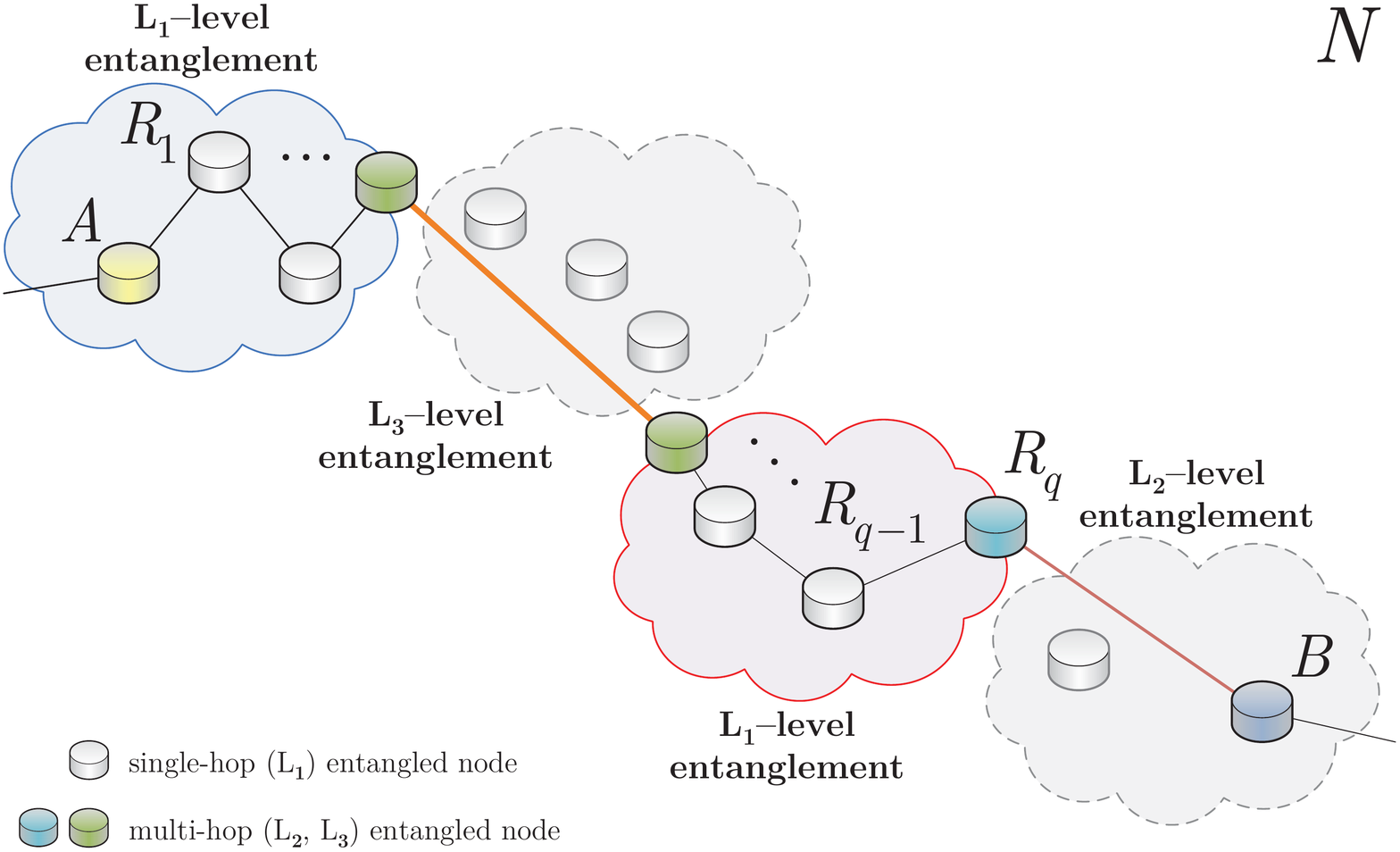}

\caption{Entangled overlay quantum network $N=\left(V,E\right)$ with heterogeneous entanglement levels. The network consists of single-hop entangled (gray) nodes with $\text{L}_1$-level entanglement connection, and multi-hop entangled (blue, green) nodes with $\text{L}_2$ and $\text{L}_3$-level entangled links. An $\text{L}_l$-level, $l=1,2,3$, entangled link between nodes $x,y\in V$ is established with probability ${\Pr }_{\text{L}_l}\left(E\left(x,y\right)\right)$. The overlay network consists of $q$ quantum repeater nodes $R_i\in V$, $i=1,\dots ,q$ between the transmitter ($A$) and the receiver ($B$) nodes. The $\text{L}_l$-level entangled nodes consist of $d{\left(x,y\right)}_{\text{L}_l}-1$ intermediate quantum nodes, as depicted by the dashed lines.}
 \label{fig1}
 \end{center}
\end{figure*}
\end{center}

An entangled overlay quantum network $N$ is illustrated in \fref{fig1}. The network consists of single-hop entangled nodes (depicted by gray nodes) and multi-hop entangled nodes (depicted by blue and green nodes) connected by edges. The single-hop entangled nodes are directly connected through an $\text{L}_1$-level entanglement, while the multi-hop entangled nodes communicate with each other through $\text{L}_2$ and $\text{L}_3$-level entanglement. Each entanglement level exists with a given probability.

\subsection{Problem Setting and Available Resources}
The proposed network model handles the quantum nodes and the quantum links in an abstract level. The quantum nodes are represented by nodes, while the quantum links are modeled by edges in a graph. The quantum links are formulated by bipartite or multipartite entangled states between the quantum nodes. The entangled quantum links are built-up by the physical-layer procedures and resource allocation mechanisms of entanglement distribution \cite{ref1,ref2, ref3, ref4, ref5, ref6, ref7, ref8}, such as entanglement purification, entanglement swapping, and quantum error correction \cite{ref15, ref16, ref17, ref18, ref19, ref20, ref21, ref22, ref23, ref24, ref25, ref26, ref27, ref28,ref42,ref43,ref44,ref45}. In the system model, if a new entangled connection is required to establish a shortest path, these physical-layer procedures are called in the background. Note, that the quantum nodes also utilize classical links to perform some auxiliary communications (see \sref{classical}) connected to the mechanisms of quantum-layer such as entanglement distribution and node selection, distribution of measurement information and statistical information between the neighboring nodes, and other related information connected to the decentralized routing mechanism. The aim of the proposed system model is to handle these procedures in an abstracted background layer that allows us to focus only on the path selection problem.

\subsubsection{Probability of Entanglement and Entanglement Fidelity}
The $F$ fidelity of entanglement \cite{ref1,ref2,ref6, ref51,ref52,ref53} at a particular density matrix $\sigma $ between nodes $x$ and $y$ is defined as $F=\langle \Psi | \sigma |\Psi \rangle $, where ${| \Psi \rangle} $ refers to the entangled system subject to be established between $x$ and $y$. Let's assume that $\sigma $ is the density matrix associated with a particular link $E\left(x,y\right)$ as $\sigma =\sum _{i}p_{i}  \rho _{i} =\sum _{i}p_{i}  {\left| \psi _{i}  \right\rangle} {\left\langle \psi _{i}  \right|} $, thus the $F_{E\left(x,y\right)} $ entanglement fidelity between nodes $x$ and $y$ is as
\begin{equation} \label{eqn1} 
{{F}_{E( x,y)}}=\langle \Psi  | \sigma  |\Psi  \rangle =\sum\limits_{i}{{{p}_{i}}{{| \langle  \Psi  | {{\psi }_{i}} \rangle |}^{2}}}. 
\end{equation}
Independent of the $\Pr _{{\rm L}_{l} } \left(E\left(x,y\right)\right)$ probability of entanglement between the nodes, in the proposed routing method each $E\left(x,y\right)$ link can be also associated with a particular entanglement fidelity [see \eqref{eqn1}]. As a corollary, $F_{E\left(x,y\right)} $ can also be selected as a routing metric in our model to find the shortest path in the quantum network. However, the $\Pr _{{\rm L}_{l} } \left(E\left(x,y\right)\right)$ probability of entanglement represents a more generalized metric that includes the effects of link noise, the effects of entanglement purification and entanglement swapping, error-correction, and disturbances of the physical environment.

Note that recent approaches to quantum networks employ quantum error correction in addition to, or instead of, entanglement purification \cite{ref46}; therefore, in these networks the effect of entanglement purification on the entanglement probability is weighted by a particular weight coefficient $\omega $, $\omega <1$, or neglected, $\omega =0$.

\subsection{Base-Graph Construction}
The base-graph \cite{ref9, ref10, ref11, ref14} of an entangled quantum network $N$ is determined as follows. Let $V$ be the set of nodes of the overlay quantum network. Then let $G^k$ be the $k$-dimensional, $n$-sized finite square-lattice base-graph \cite{ref1, ref9, ref10, ref12, ref13, ref14}, with position $\phi \left(x\right)$ assigned to an overlay quantum network node $x\in V$, where $\phi :V\to G^k$ is a mapping function which achieves the mapping from $V$ onto $G^k$ \cite{ref10}.

Specifically, for two network nodes $x,y\in V$, the L1 metric in $G^k$ is denoted by $d\left(\phi \left(x\right),\phi \left(y\right)\right)$, $\phi \left(x\right)=\left(j,k\right)$, $\phi \left(y\right)=\left(m,o\right)$ and is defined as 
\begin{equation} \label{eq2} 
d\left(\left(j,k\right),\left(m,o\right)\right)=\left|m-j\right|+\left|o-k\right|. 
\end{equation}

\subsubsection{Connection Probabilities }
The $G^k$ base-graph contains all entangled contacts of all $x\in V$. The probability that $\phi \left(x\right)$ and $\phi \left(y\right)$ are connected through an $\text{L}_l$-level entanglement in $G^k$ is 
\begin{equation} \label{eq3} 
p\left(\phi \left(x\right),\phi \left(y\right)\right)=\frac{d{\left(\phi \left(x\right),\phi \left(y\right)\right)}^{-k}}{H_n}+c_{\phi \left(x\right),\phi \left(y\right)}, 
\end{equation} 
 where 
\begin{equation} \label{eq4} 
H_n=\sum_z{}d\left(\phi \left(x\right),\phi \left(z\right)\right) 
\end{equation} 
 is a normalizing term \cite{ref9, ref10}, which is taken over all entangled contacts of node $\phi \left(x\right)$ in $G^k$, while $c_{\phi \left(x\right),\phi \left(y\right)}$ is a constant defined as 
\begin{equation} \label{eq5} 
c_{\phi \left(x\right),\phi \left(y\right)}={\Pr }_{\text{L}_l}\left(E\left(x,y\right)\right)-\frac{d{\left(\phi \left(x\right),\phi \left(y\right)\right)}^{-k}}{H_n}, 
\end{equation} 
 where ${\Pr }_{\text{L}_l}\left(E\left(x,y\right)\right)$ is the probability that nodes $x,y\in V$ are connected through an $\text{L}_l$-level entanglement in the overlay quantum network $N$.

For an $\text{L}_l$-level entanglement between $\phi \left(x\right)$ and $\phi \left(y\right)$, $d\left(\phi \left(x\right),\phi \left(y\right)\right)$ in $G^k$ is evaluated as 
\begin{equation} \label{eq6} 
d\left(\phi \left(x\right),\phi \left(y\right)\right)=2^{l-1}. 
\end{equation} 
 Our idea is that the ${\Pr }_{\text{L}_l}\left(E\left(x_i,y_i\right)\right)$ probability of an $\text{L}_i$-level entanglement connection between nodes $x_i,y_i\in V$ in the entangled overlay quantum network $N$ can be rephrased directly by the probability of $p\left(\phi \left(x_i\right),\phi \left(y_i\right)\right)$ in the $k$-dimensional base-graph $G^k$ via the following distance connection: 
\begin{equation} \label{eq7} 
d\left(\phi \left(x_i\right),\phi \left(y_i\right)\right)=d{\left(x_i,y_i\right)}_{\text{L}_l}=2^{l-1}. 
\end{equation} 
 Between the $\phi \left(\cdot \right)$ configuration of positions of the quantum nodes in $G^k$ and the set $E$ of the $m$ edges of the overlay network $V$, the following conditional probability can be defined: 
\begin{equation} \label{eq8} 
\Pr \left(\left.E\right|\phi \right)=\prod^m_{E_i=1}{}\frac{d{\left(\phi \left(x_i\right),\phi \left(y_i\right)\right)}^{-k}}{H_n}+c_{\phi \left(x_i\right),\phi \left(y_i\right)}, 
\end{equation} 
 where $x_i,y_i\in V$ are the quantum nodes connected via an entangled link $E_i$ in the overlay network $N$.

Thus, the mapping $V\to G^k$ holds the connectivity of $V$ via the unique position configurations $\phi \left(x_i\right),\phi \left(y_i\right)$ of the overlay nodes such that the probability of an edge in $G^k$ depends only on the distance $d\left(\phi \left(x_i\right),\phi \left(y_i\right)\right)$ between $\phi \left(x_i\right),\phi \left(y_i\right)$ and the corresponding ${\Pr }_{\text{L}_i}\left(E\left(x_i,y_i\right)\right)$ in $N$.

As follows from \eqref{eq8}, to maximize $\Pr \left(\left.E\right|\phi \right)$ we have to determine those base-graph $\phi \left(x_i\right)\in G^k$ assignments for all i of overlay nodes $x_i\in V$ that minimize the product of the $d\left(\cdot \right)$ distances in the base-graph $G^k$.
 
\subsubsection{Quantum Nodes and Entangled Links onto a Base-Graph}
In particular, using stochastic optimization at a given set of $m$ edges $E$ of the overlay quantum network $N$, finding the positions $\phi \left(x_i\right),\phi \left(y_i\right)$, $i=1,\dots ,m$ in $G^k$ can be approached straightforwardly by Bayes' rule as 
\begin{equation} \label{eq9} 
\Pr \left(\left.\phi \right|E\right)=\frac{\Pr \left(\left.E\right|\phi \right)\Pr \left(\phi \right)}{\Pr \left(E\right)}, 
\end{equation} 
 which characterizes the a posteriori distribution of configuration $\phi $ at a given set $E$. Therefore, the $\phi :V\to G^k$ mapping function which maximizes $\Pr \left(\left.\phi \right|E\right)$ can be determined via a statistical estimation.

For a candidate distribution $\Pr \left(\phi \right)$, $\Pr \left(\left.\phi \right|E\right)$ can be rewritten without loss of generality as 
\begin{equation} \label{eq10} 
\Pr \left(\left.\phi \right|E\right)=\frac{\Pr \left(\left.E\right|\phi \right)\Pr \left(\phi \right)}{\int\limits_{\phi }^{}\Pr \left(\left.E\right|\phi \right)\Pr \left(\phi \right)d\phi }, 
\end{equation} 
 which clearly reveals that the determination of \eqref{eq10}, specifically the computation 
\begin{equation} \label{eq11} 
\int\limits_{\phi }^{}\Pr \left(\left.E\right|\phi \right)\Pr \left(\phi \right)d\phi , 
\end{equation} 
 is also hard \cite{ref10, ref11, ref14}. To solve the problem, Markov chain--based techniques \cite{ref10} can be utilized, allowing us to generate samples of $\phi $ that conform to a given $\Pr \left(\phi \right)$ candidate distribution \cite{ref10} (see \sref{chain}); this is convenient since we can determine the denominator of \eqref{eq10}. These techniques require the definition of a proposal density function to stabilize the resulting Markov chain. This stabilization is required to achieve \eqref{eq10} via the chain through a sequence of states. A proposal density function $q\left(\left.r\right|s\right)$ proposes a next state $s^*$ given a state $s_i$.

On the other hand, the stabilization procedure also requires the swapping of position information $\phi \left(x_i\right)$ and $\phi \left(y_i\right)$ between any two nodes $\phi \left(x_i\right),\phi \left(y_i\right)\in G^k$ subject to some constraints. The swapping operation between two nodes does not change the physical-level connections. However, assuming a classical communication channel for this purpose, the swapping would lead to serious security issues \cite{ref10, ref14}.

\subsubsection{Swapping by Quantum Teleportation}
\label{swap}
As we prove here, by utilizing entangled links between nodes, our solution requires no transmission of information $\phi \left(x_i\right)$ and $\phi \left(y_i\right)$ between the nodes $x_i,y_i\in V$ of the overlay network for stabilization. Particularly, our stabilization procedure uses quantum teleportation between nodes, which does not require transmission of any routing-related information in the network, as follows.

Let's assume that quantum nodes $x_i,y_i\in V$ are selected for swapping from the entangled overlay network $N$, associated with $G^k$ position information $\phi \left(x_i\right)$ and $\phi \left(y_i\right)$. Let $u_j$ refer to the $j^{th}$ neighbor quantum node of $x_i$, $\left\{x_i,u_j\right\}\in E$ with position $\phi \left(u_j\right)\in G^k$, and let $v_j$ identify the $j^{th}$ neighbor quantum node of $y_i$, $\left\{y_i,v_j\right\}\in E$ with position $\phi \left(v_j\right)\in G^k$. In the first phase, all neighbor nodes of $x_i,y_i$ locally prepare the quantum systems $\left|\phi \left(u_j\right)\right\rangle $ and $\left|\phi \left(v_j\right)\right\rangle $. Using the $\text{L}_l$-level entangled links between $u_j$ and $x_i$, $v_j$ and $y_i$, all neighbor quantum nodes teleport their local quantum system to $x_i$ and $y_i$. This is possible since all nodes of $V$ are connected through an $\text{L}_l$-level entanglement in $N$, and therefore, an arbitrary neighbor node is at least connected through an $\text{L}_1$-level (direct) entanglement.

Specifically, for $\forall j$, the neighbor node $u_j$ teleports $\left|\phi \left(u_j\right)\right\rangle $ to $x_i$, while all $v_j$ teleports $\left|\phi \left(v_j\right)\right\rangle $ to $y_i$, respectively. In the next step, for $\forall j$ the nodes $x_i$ and $y_i$ measure their states $\left|\phi\left(u_j\right)\right\rangle $ and $\left|\phi \left(v_j\right)\right\rangle $ via a local measurement $M$, which yields 
\begin{equation} \label{eq12} 
M\left|\phi \left(u_j\right)\right\rangle =\phi \left(u_j\right) 
\end{equation} 
 and 
\begin{equation} \label{eq13} 
M\left|\phi \left(v_j\right)\right\rangle =\phi \left(v_j\right). 
\end{equation} 
 Using the results of the local measurements, the two nodes $x_i$ and $y_i$ determine the following quantities: 
\begin{equation} \label{eq14} 
\begin{split}
&\zeta \left(x_i,y_i\right)\\&=\prod_{\left\{x_i,u_j\right\}\in E}{}\left(\phi \left(x_i\right)-\phi \left(u_j\right)\right)\prod_{\left\{y_i,v_j\right\}\in E}{}\left(\phi \left(y_i\right)-\phi \left(v_j\right)\right), 
\end{split}
\end{equation} 
 and 
\begin{equation} \label{eq15} 
\begin{split}
&\Phi \left(x_i,y_i\right)\\&=\prod_{\left\{x_i,u_j\right\}\in E}{}\left(\phi \left(y_i\right)-\phi \left(u_j\right)\right)\prod_{\left\{y_i,v_j\right\}\in E}{}\left(\phi \left(x_i\right)-\phi \left(v_j\right)\right). 
\end{split}
\end{equation} 
 In the final step, the two nodes $x_i$ and $y_i$ make a decision regarding their location information swapping.

Particularly, if 
\begin{equation} \label{eq16} 
\zeta \left(x_i,y_i\right)\ge \Phi \left(x_i,y_i\right), 
\end{equation} 
 then nodes $x_i,y_i$ perform the swapping operation, which yields 
\begin{equation} \label{eq17} 
M\left|\phi \left(y_i\right)\right\rangle \equiv \phi \left(x_i\right) 
\end{equation} 
 at $x_i$, and 
\begin{equation} \label{eq18} 
M\left|\phi \left(x_i\right)\right\rangle \equiv \phi \left(y_i\right) 
\end{equation} 
 at $y_i$, with unit probability 
\begin{equation} \label{eq19} 
p_{swap}\left(\phi \left(x_i\right),\phi \left(y_i\right)\right)=1. 
\end{equation} 
 If 
\begin{equation} \label{eq20} 
\zeta \left(x_i,y_i\right)<\Phi \left(x_i,y_i\right), 
\end{equation} 
 then nodes $x_i,y_i$ swap their position information only with probability 
\begin{equation} \label{eq21} 
{{p}_{swap}}\left( \phi \left( {{x}_{i}} \right),\phi \left( {{y}_{i}} \right) \right)=\frac{\zeta \left( {{x}_{i}},{{y}_{i}} \right)}{\Phi \left( {{x}_{i}},{{y}_{i}} \right)}, 
\end{equation} 
 which is also a possible scenario if the nodes $x_i,y_i$ are uniformly selected at random \cite{ref11}.

Applying the swapping procedure for all node pairs of $V$ provably stabilizes the chain since it leads to the convergence of the $\phi \left(\cdot \right)$ positions to a state which allows us to perform efficient decentralized routing in the $G^k$ base-graph, using the L1 metric. 
 
\paragraph{Markov Chain}
\label{chain}
  The Markov chain for the base-graph construction is defined as follows. Let ${\phi }_2$ be the $x_i,y_i$-swap of ${\phi }_1$, such that ${\phi }_1\left(x_i\right)={\phi }_2\left(y_i\right)$, ${\phi }_1\left(y_i\right)={\phi }_2\left(x_i\right)$, and ${\phi }_1\left(z_i\right)={\phi }_2\left(z_i\right)$ for all $z_i\ne x_i,y_i$ \cite{ref10, ref14}. Then let the Markov chain defined by transition matrix $T\left({\phi }_1,{\phi }_2\right)$, as $T\left({\phi }_1,{\phi }_2\right)=\Omega \left({\phi }_1,{\phi }_2\right)\varepsilon \left({\phi }_1,{\phi }_2\right)$, where ${\phi }_1\ne {\phi }_2$. If ${\phi }_2$ is the $x_i,y_i$-swap of ${\phi }_1$, then $\Omega \left( {{\phi }_{1}},{{\phi }_{2}} \right)={1}/{\left( n+\left( \begin{smallmatrix}
   n  \\
   2  \\
\end{smallmatrix} \right) \right)}$, and $\Omega \left(\phi \left(x_i\right),\phi \left(y_i\right)\right)=0$ otherwise \cite{ref10}. The term $\varepsilon \left({\phi }_1,{\phi }_2\right)$ is defined as 
\begin{equation} \label{eq22}
\begin{split} 
&\varepsilon \left({\phi }_1,{\phi }_2\right)\\&=\min \left(1,\prod_{E_i\in E\left(x\vee y\right)}{}\frac{d{\left({\phi }_1\left(x_i\right),{\phi }_1\left(y_i\right)\right)}^k+c_{{\phi }_1\left(x_i\right),{\phi }_1\left(y_i\right)}}{d{\left({\phi }_2\left(x_i\right),{\phi }_2\left(y_i\right)\right)}^k+c_{{\phi }_2\left(x_i\right),{\phi }_2\left(y_i\right)}}\right), 
\end{split} 
\end{equation} 
 where $E\left(x\vee y\right)$ refers to the edges connected to $x\in V$ or $y\in V$; therefore, $\varepsilon \left({\phi }_1,{\phi }_2\right)$ can be determined via each node by only its local edge information.

As one can readily check, the chain with $T\left({\phi }_1,{\phi }_2\right)$ has $\Pr \left(\left.\phi \right|E\right)$ [see \eqref{eq9}] as its stationary distribution.

\subsection{Next-Generation Repeaters}
\label{nextgen}
The result in \eqref{eq1} reflects the characteristic of the entanglement distribution mechanism of the doubling-architecture \cite{ref1,ref3,ref7,ref8}. On the other hand, the proposed routing method can also be extended to third-generation quantum repeater quantum networks \cite{ref46} that do not necessarily involve the establishment of long-distance entangled links. In this terminology, \eqref{eq1} identifies the $d\left(x,y\right)$ hop-distance between quantum nodes $x$ and $y$ in the network, without the utilization of entangled links and the level characteristics of the doubling-architecture. As a corollary, for a third-generation quantum repeater network setting the level ${\rm L}_{l} $ of a $E\left(x,y\right)$ link refers directly to the hop-distance, i.e., $l$ is set as $l=d\left(x,y\right)$. Therefore, the proposed routing method remains directly applicable in next-generation quantum repeater networks, since the links between the quantum nodes can also be associated with a particular link probability $\Pr _{{\rm L}_{l} } \left(E\left(x,y\right)\right)$. Note the swapping mechanism of \sref{swap} for these networking scenarios can be established via secure quantum communications.

\subsection{Classical Communications in the Quantum Network}
\label{classical}
The proposed method also utilizes some classical communications to perform the decentralized routing to find a shortest path in the quantum network. Without loss of generality, a classical communication phase consists of the selection of the quantum nodes, local communications between the neighboring quantum nodes, distribution of measurement information between the neighboring nodes, and sharing of statistical information regarding the entangled links. The locally distributed measurement information consists of the measurement results of the quantum teleportation procedure (see \sref{swap}), and other measurements results connected to the entanglement distribution mechanism (e.g., entanglement purification, entanglement swapping, quantum error correction, etc) in the quantum network.

\section{Decentralized Routing in the Base-Graph}
\label{sec3}
The routing in the $k$-dimensional base-graph $G^k$ is performed via a decentralized algorithm $\mathcal{A}$ as follows. After we have determined the base-graph $G^k$ of the entangled overlay quantum network $N$, we can apply the L1 metric to find the shortest paths. Since the probability that two arbitrary entangled nodes $\phi \left(x\right),\phi \left(y\right)$ are connected through an $\text{L}_l$-level entanglement is $p\left(\phi \left(x\right),\phi \left(y\right)\right)$ [see \eqref{eq3}], this probability distribution associated with the entangled connectivity in $G^k$ allows us to achieve efficient decentralized routing via $\mathcal{A}$ in the base-graph.

Using the L1 distance function, a greedy routing (which always selects a neighbor node closest to the destination node in terms of $G^k$ distance function $d$ and does not select the same node twice) can be straightforwardly performed in $G^k$ to find the shortest path from any quantum node to any other quantum node, in 
\begin{equation} \label{eq23} 
\mathcal{O}{\left(\log n\right)}^2 
\end{equation} 
 steps on average (see \sref{subsec3}), where $n$ is the size of the network of $G^k$.

Note that the nodes know only their local links (neighbor nodes) and the target position. It also allows us to avoid dead-end nodes (where the routing would stop) by some constraints on the degrees of the nodes, which can be directly satisfied through the settings of the overlay quantum network.

The decentralized algorithm $\mathcal{A}$ in the $k$-dimensional $n$-sized base-graph $G^k$ is characterized by the following diameter bounds.

In our setting, the $D\left(G^k\right)$ diameter of $G^k$ refers to the maximum value of the shortest path (total number of edges on a path) between any pair of mapped nodes in $G^k$.

Then, for the $D\left(\mathcal{A}\right)$ minimal number of steps required by $\mathcal{A}$ follows that 
\begin{equation} \label{eq24} 
D\left(\mathcal{A}\right)\ge D\left(G^k\right). 
\end{equation} 
 We show that for any $G^k$ with $p\left(\phi \left(x\right),\phi \left(y\right)\right)$ [see \eqref{eq3}] probability for the entangled links between an arbitrary $\phi \left(x\right),\phi \left(y\right)\in G^k$, the relation 
\begin{equation} \label{eq25} 
D\left(\mathcal{A}\right)\le \mathcal{O}{\left(\log n\right)}^2 
\end{equation} 
 holds.

In \sref{subsec3} we prove that for any $G^k$, the relation of \eqref{eq25} holds.

\begin{figure}[!h]
 \begin{center}
 	 \includegraphics[angle = 0,width=1\linewidth]{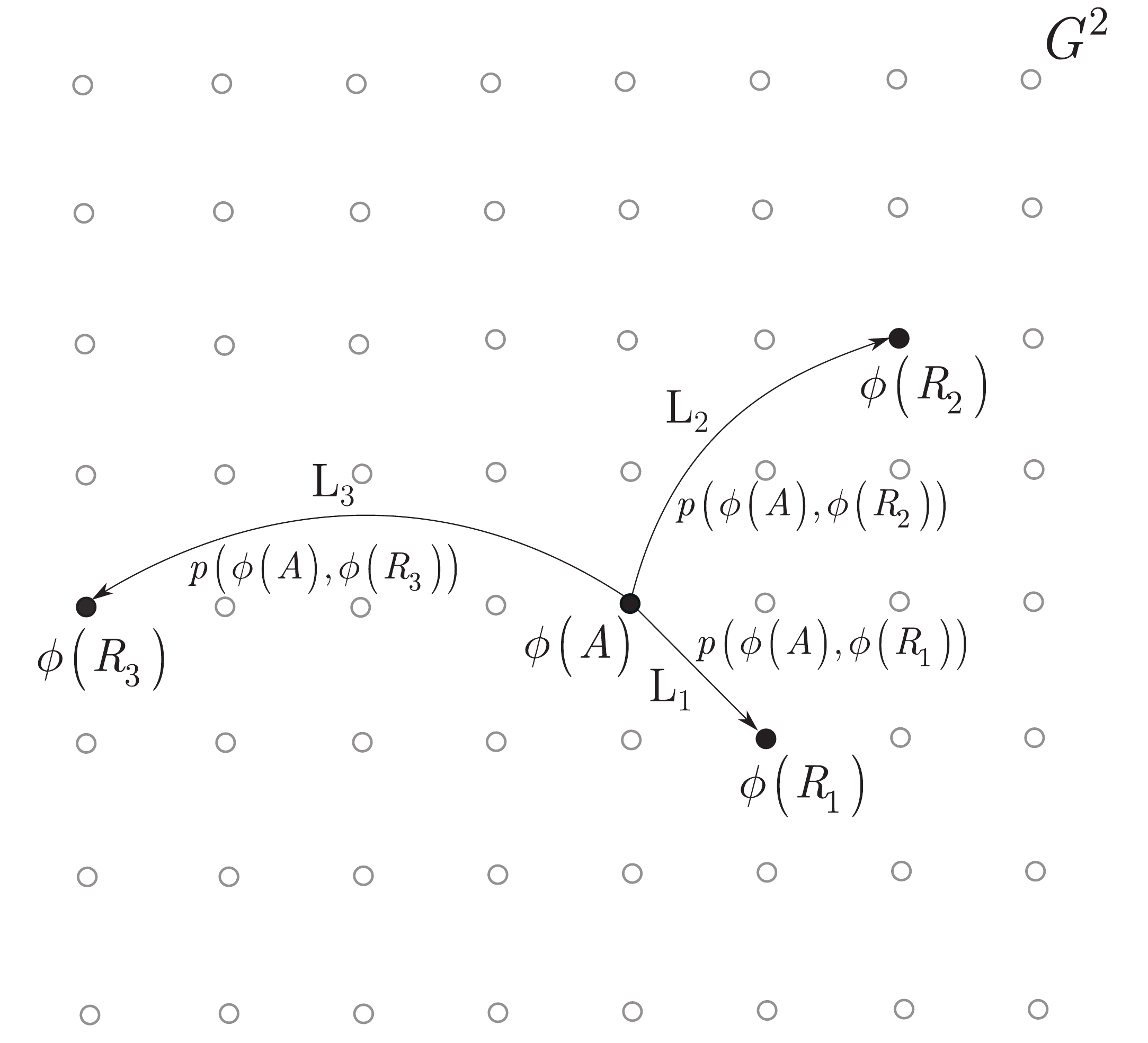}

\caption{$G^2$ base-graph of an overlay quantum network $N$, with entangled nodes $\phi \left(A\right)$, $\phi \left(R_i\right)$, $i=1,2,3$, where $A\in V$ is a transmitter node in the overlay quantum network $N$, while $R_i\in V$ are quantum repeater nodes in $N$. In $N$, nodes $A$ and $R_1$ are connected through $\text{L}_1$-level entanglement with probability ${\Pr }_{\text{L}_1}\left(A,R_1\right)$, nodes $A$ and $R_2$ are connected via $\text{L}_2$-level entanglement with probability ${\Pr }_{\text{L}_2}\left(A,R_2\right)$, while $A$ and $R_3$ have an $\text{L}_3$-level entanglement connection with probability ${\Pr }_{\text{L}_3}\left(A,R_3\right)$. The probability that nodes are connected in $G^2$ is $p\left(\phi \left(A\right),\phi \left(R_1\right)\right)$, $p\left(\phi \left(A\right),\phi \left(R_2\right)\right)$, and $p\left(\phi \left(A\right),\phi \left(R_3\right)\right)$.}
 \label{fig2}
\end{center}
\end{figure}

In \fref{fig2}, a $G^k$, $k=2$ dimensional base-graph is depicted with entangled nodes $\phi \left(A\right)\in G^2$, $\phi \left(R_i\right)\in G^2$, $i=1,2,3$, where $A\in V$ is a transmitter node in the overlay quantum network $V$, while $R_i\in V$ are quantum repeater nodes in $N$. The nodes are connected through an $\text{L}_i$-level entanglement in $N$ with probability ${\Pr }_{\text{L}_i}\left(A,R_i\right)$. In the base-graph $G^2$, the mapped nodes $\phi \left(A\right)$, $\phi \left(R_i\right)$ are connected with probability $p\left(\phi \left(A\right),\phi \left(R_i\right)\right)=\frac{d{\left(\phi \left(A\right),\phi \left(R_i\right)\right)}^{-2}}{\sum_z{}d{\left(\phi \left(A\right),\phi \left(R_z\right)\right)}^{-2}}+c_{\phi \left(A\right),\phi \left(R_i\right)},$ where $d\left(\phi \left(A\right),\phi \left(R_i\right)\right)=2^{i-1}$.

\subsection{Routing Complexity}
\label{subsec3}
  In this section we prove that for our decentralized algorithm $\mathcal{A}$, for an arbitrary $k$-dimensional $n$-size base-graph $G^k$, the relation of 
\begin{equation} \label{eq27} 
D\left(\mathcal{A}\right)\le \mathcal{O}{\left(\log n\right)}^2 
\end{equation} 
 holds.

Utilizing the tessellation of $B_n$ for $m$ times results in end squares with side length $n^{{\gamma }^m}$, for which situation $m$ events, $A_1,\dots ,A_m$, exist \cite{ref12}. In this case, the resulting bound on the diameter is 
\begin{equation} \label{eq28} 
D\left(G^2\right)\le 2^{m+2}n^{{\gamma }^m}. 
\end{equation} 
 It can be verified that 
\begin{equation} \label{eq29} 
m=\left(\log \log n-\log \log \log n+\log \left(4\gamma -k\right)-\log K\right)/\log {\gamma }^{-1}, 
\end{equation} 
 where $K$ is a constant \cite{ref10, ref11, ref12}, and 
\begin{equation} \label{eq30} 
{\gamma }^m=\frac{K\log \log n}{\left(4\gamma -k\right)\log n}, 
\end{equation} 
 threfore, the diameter bound is as 
\begin{equation} \label{eq31} 
D\left(G^2\right)\le {\left(\log n\right)}^C, 
\end{equation} 
 for some constant $C>0$, which leads to 
\begin{equation} \label{eq32} 
\underset{n\to \infty }{\mathop{\lim }}\,\Pr \left( D\left( {{G}^{2}} \right)\le {{\left( \log n \right)}^{C}} \right)=1.
\end{equation} 
 Note, that the probability that an event $A_i$ occurs (i.e., there is no edge between the $n^{{\gamma }^i}$ side subsquares) is bounded by 
\begin{equation} \label{eq33} 
\Pr \left( {{A}_{i}} \right)\le {{n}^{4}}{{e}^{-Z{{n}^{{{\gamma }^{i-1}}}}\left( 4\gamma -k \right)}}, 
\end{equation} 
 where $Z>0$ is a constant, while ${{n}^{{{\gamma }^{i-1}}}}$ refers to the large subsquare which is tessellated by the ${{n}^{{{\gamma }^{i}}}}$ side sub-subsquares, respectively. Thus, 
\begin{equation} \label{eq34} 
\Pr \left( {{A}_{1}}\wedge \ldots \wedge {{A}_{m}} \right)\le m{{n}^{4}}{{e}^{-Z{{n}^{{{\gamma }^{m}}}}\left( 4\gamma -k \right)}}. 
\end{equation} 
 To verify the upper bound \eqref{eq27}, we use the fact that for any $\phi \left(x\right)\in G^2$, by theory 
\begin{equation} \label{eq35} 
\sum_{\phi \left(y\right)\in G^2,\phi \left(y\right)\ne \phi \left(x\right)}{}{\left(d\left(\phi \left(x\right)-\phi \left(y\right)\right)\right)}^{-2}\le 4\log \left(6n\right) 
\end{equation} 
 from which the probability 
\begin{equation} \label{eq36} 
\Pr \left(\left.\phi \left(y\right)\right|\phi \left(x\right)\right) 
\end{equation} 
 that from node $\phi \left(x\right)$ a given $\phi \left(y\right)$ is selected is lower bounded by 
\begin{equation} \label{eq37} 
\Pr \left(\left.\phi \left(y\right)\right|\phi \left(x\right)\right)\ge \frac{d{\left(\phi \left(x\right)-\phi \left(y\right)\right)}^{-2}}{4\log \left(6n\right)}. 
\end{equation} 
 Then let $e_j$, be an event that from node $\phi \left(x\right)$ a set ${\mathcal{S}}_j$ of nodes can be selected by $\mathcal{A}$, where 
\begin{equation} \label{eq38} 
j\in \left[\log \log n,\log n\right], 
\end{equation} 
 such that ${\mathcal{S}}_j$ are within L1 distance $2^j$ from the target node $\phi \left(B\right)$.

In set ${\mathcal{S}}_j$, each node is within the L1 distance 
\begin{equation} \label{eq39} 
2^{j+1}+2^j<2^{j+2} 
\end{equation} 
 of $\phi \left(x\right)$. After some calculations \cite{ref9, ref12}, the probability that an event $e_j$ occurs is 
\begin{equation} \label{eq40} 
\Pr \left(e_j\right)\ge \frac{1}{64\log \left(6n\right)}. 
\end{equation} 
 Therefore, if the current node is $\phi \left(x\right)$, and 
\begin{equation} \label{eq41} 
2^j<d\left(\phi \left(x\right),\phi \left(B\right)\right)\le 2^{j+1} 
\end{equation} 
 holds for the L1 distance, then the number of steps are upper bounded by the mean $E\left(X_j\right)$ of an geometric random variable $X_j$, 
\begin{equation} \label{eq42} 
E\left(X_j\right)=\frac{1}{\Pr \left(e_j\right)}=\mathcal{O}\left(\log n\right). 
\end{equation} 
 Since the number of such events is maximized in $\log n$, it immediately follows that the total number of steps in $G^2$ is on average at most $\mathcal{O}{\left(\log n\right)}^2$, thus 
\begin{equation} \label{eq43} 
D\left(\mathcal{A}\right)\le \log n\frac{1}{\Pr \left(e_j\right)}=\mathcal{O}{\left(\log n\right)}^2, 
\end{equation} 
 which holds for an arbitrary, $k$-dimensional $n$-size base-graph $G^k$. 

\subsection{Implementation}
Since the proposed method requires no additional physical apparatus in an experimental quantum networking scenario, the algorithm in a stationary quantum node can be implemented by standard photonics devices, quantum memories, optical cavities and other fundamental physical devices currently in practical use in experimental quantum networking \cite{ref15,ref16,ref17,ref18,ref19,ref20,ref21,ref22,ref23,ref24,ref25,ref26,ref27}. The quantum transmission and the auxiliary classical communications between the nodes can be realized via standard links (i.e., optical fibers, wireless optical channels, free-space quantum channels, etc), and by the application of fundamental quantum transmission protocols of quantum networks \cite{ref28,ref29,ref30,ref31,ref32,ref33,ref34,ref35,ref36,ref37,ref38,ref39,ref40}. 

\subsubsection{Practical Benefits}
The practical benefits of this work in the context of an actual quantum network are as follows. Since the proposed routing has a low-complexity, it allows resource savings in the quantum nodes. Both the overall storage time of the quantum states in the local quantum memories of the quantum nodes, and the number of auxiliary communications and internal computational steps related to the path determination in the nodes can be minimized. As a corollary, the proposed decentralized routing method has a minimal overall delay in the quantum network that has a crucial significance in an experimental quantum network setting.  

\section{Diameter Bounds}
\label{sec4}
Here we derive the diameter bounds for a $k=2$ dimensional $n$-size base-graph $G^2$. The results can be extended for arbitrary dimensions.

Let $B_n$ be a box of size $n\times n$ that contains $G^2$. Let $S_i$ be a subsquare of $B_n$ of side length ${{n}^{\gamma }}$, where 
\begin{equation} \label{eq44} 
{k}/{4}\;<\gamma <1, 
\end{equation} 
 and let us subdivide each $S_i$ into smaller sub-subsquares $S_{ik}$ of side length ${{n}^{{{\gamma }^{2}}}}$ \cite{ref12}.

Let $A_1$ be the event that there exists at least two subsquares $S_i$ and $S_j$ in $B_n$ such that there is exists no edge between them. Similarly, let $A_2$ identify the event that exists at one $S_i$ in $B_n$ such that there are two sub-subsquares $S_{ik}$ in $S_i$ which are not connected by edge. In particular, assuming a $G^2$ for which $A_1$ is violated means that subsquares $S_i$ and $S_j$ are connected by at least one edge, thus without loss of generality, 
\begin{equation} \label{eq45} 
D\left(G^2\right)\le 2D_{max}\left(S_i\right)+1, 
\end{equation} 
 where $D_{max}\left(S_i\right)$ identifies the largest diameter of the subsquares of side length ${{n}^{\gamma }}$. By similar assumptions, if $A_2$ is violated then there exists an edge between at least two sub-subsquares $S_{ik}$ of any $S_i$; therefore,
\begin{equation} \label{eq46} 
D\left(G^2\right)\le 4D_{max}\left(S_{ik}\right)+3, 
\end{equation} 
 where $D_{max}\left(S_{ik}\right)$ is the largest diameter of the sub-subsquares of side length ${{n}^{{{\gamma }^{2}}}}$, respectively. As follows, in this case there exists a path of length 
\begin{equation} \label{eq47} 
D\left(G^2\right)\le 4D_{max}\left(S_{ik}\right)+3 
\end{equation} 
 in $B_n$ which connects any two mapped nodes $\phi \left(x\right),\phi \left(y\right)$ in $G^2$.

\begin{figure}[!h]
 \begin{center}
 	 \includegraphics[angle = 0,width=1\linewidth]{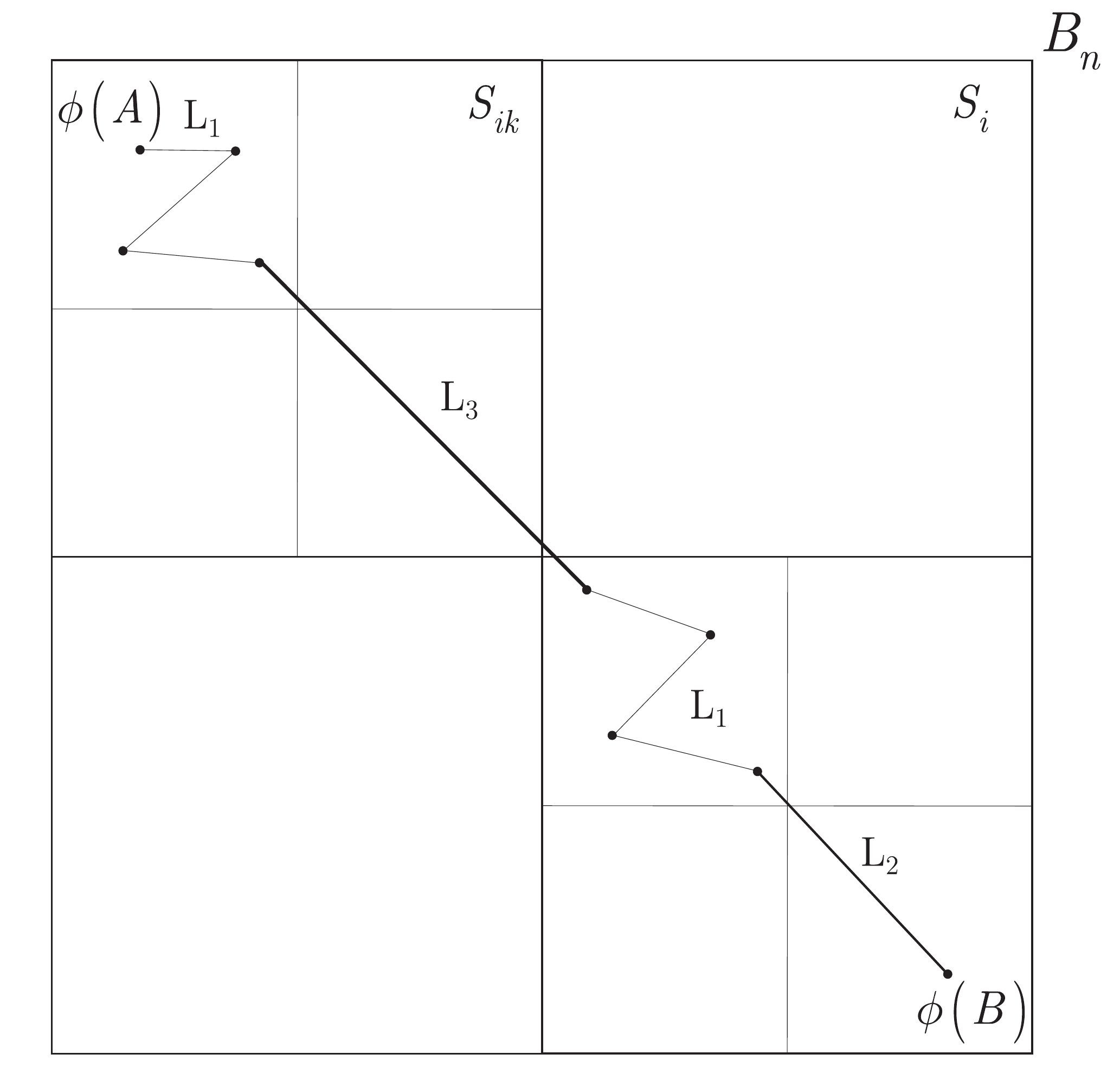}

\caption{A tessellation of ${{B}_{n}}$ of the base-graph ${{G}^{2}}$ of an overlay quantum network $N$ onto ${{n}^{\gamma }}$ side subsquare ${{S}_{i}}$, and ${{n}^{{{\gamma }^{2}}}}$ side sub-subsquare ${{S}_{ik}}$, where ${k}/{4}\;<\gamma <1$. The nodes are connected through ${{\text{L}}_{1}},{{\text{L}}_{2}}$ and ${{\text{L}}_{3}}$-level entangled links in the overlay network, with source node $A$ and target node $B$. The points between $\phi \left( A \right)$ and $\phi \left( B \right)$ refer to the repeater quantum nodes.}
\label{fig3}
\end{center}
\end{figure}

Tessellation of a base-graph $G^2$ of an overlay quantum network $N$ for which these events are violated is illustrated in \fref{fig3}. The $B_n$ box contains $G^2$, with subsquares $S_i$, and sub-subsquares $S_{ik}$. The nodes are connected through $\text{L}_1,\text{L}_2$ and $\text{L}_3$-level entanglement in $N$.

\section{Conclusions}
\label{con}
We proposed a method to perform efficient decentralized routing in the entangled networks of the quantum Internet. Our solution allows us to find the shortest path in multi-level entangled quantum networks of the quantum Internet, using only local knowledge of the nodes. We showed that the entangled network structure can be embedded onto a base-graph, keeping the probability distribution of the entangled links and allowing us to construct efficient decentralized routing. The results can be directly applied in practical quantum communications, experimental long-distance quantum key distribution, quantum repeater networks, future quantum Internet, and quantum networking scenarios. 

\acknowledgements

This work was partially supported by the National Research Development and Innovation Office of Hungary (Project No. 2017-1.2.1-NKP-2017-00001), by the Hungarian Scientific Research Fund - OTKA K-112125 and in part by the BME Artificial Intelligence FIKP grant of EMMI (BME FIKP-MI/SC).

\bibliography{report}
\bibliographystyle{unsrt}

\clearpage
\onecolumngrid
\appendix
\setcounter{table}{0}
\setcounter{figure}{0}
\setcounter{equation}{0}
\setcounter{algocf}{0}
\renewcommand{\thetable}{\Alph{section}.\arabic{table}}
\renewcommand{\thefigure}{\Alph{section}.\arabic{figure}}
\renewcommand{\theequation}{\Alph{section}.\arabic{equation}}
\renewcommand{\thealgocf}{\Alph{section}.\arabic{algocf}}

\setlength{\arrayrulewidth}{0.1mm}
\setlength{\tabcolsep}{5pt}
\renewcommand{\arraystretch}{1.5}
\section{Notations}

The notations of the manuscript are summarized in \tref{tab1}.
\begin{center}
\begin{longtable}{||l|p{3.9in}||}
\caption{Summary of notations.}
\label{tab1}
\endfirsthead
\endhead
\hline
\textit{Notation} & \textit{Description} \\ \hline 
L1 & Manhattan distance (L1 metric). \\ \hline 
$l$ & Level of entanglement.  \\ \hline 
$F$ & Fidelity of entanglement.  \\ \hline 
$\text{L}_l$ & An $l$-level entangled link. For an $\text{L}_l$ link, the hop-distance is $2^{l-1}$. \\ \hline 
$d{\left(x,y\right)}_{\text{L}_l}$ & Hop-distance of an $l$-level entangled link between nodes $x$ and $y$.  \\ \hline 
$\text{L}_1$ & $\text{L}_1$-level (direct) entanglement,  $d{\left(x,y\right)}_{\text{L}_1}=2^0=1$. \\ \hline 
$\text{L}_2$ & $\text{L}_2$-level entanglement, $d{\left(x,y\right)}_{\text{L}_2}=2^1=2$. \\ \hline 
$\text{L}_3$ & $\text{L}_3$-level entanglement, $d{\left(x,y\right)}_{\text{L}_3}=2^2=4$. \\ \hline 
$E\left(x,y\right)$ & An edge between quantum nodes $x$ and $y$, refers to an $\text{L}_l$-level entangled link. \\ \hline 
${\Pr }_{\text{L}_l}\left(E\left(x,y\right)\right)$ & Probability of existence of an entangled link $E\left(x,y\right)$, $0<{\Pr }_{\text{L}_l}\left(E\left(x,y\right)\right)\le 1$. \\ \hline 
$N$ & Overlay quantum network, $N=\left(V,E\right)$, where $V$ is the set of nodes, $E$ is the set of edges. \\ \hline 
$V$ & Set of nodes of $N$. \\ \hline 
$E$ & Set of edges of $N$. \\ \hline 
$G^k$ & An $n$-size, $k$-dimensional base-graph. \\ \hline 
$n$ & Size of base-graph $G^k$. \\ \hline 
$k$ & Dimension of base-graph $G^k$. \\ \hline 
$A$ & Transmitter node, $A\in V$. \\ \hline 
$B$ & Receiver node, $B\in V$. \\ \hline 
$R_i$ & A repeater node in $V$, $R_i\in V$. \\ \hline 
$E_j$ & Identifies an $\text{L}_l$-level entanglement, $l=1,\dots ,r$, between quantum nodes $x_j$ and $y_j$. \\ \hline 
$E=\left\{E_j\right\}$ & Let $E=\left\{E_j\right\}$, $j=1,\dots ,m$ refer to a set of edges between the nodes of $V$. \\ \hline 
$\phi \left(x\right)$ & Position assigned to an overlay quantum network node $x\in V$ in a $k$-dimensional, $n$-sized finite square-lattice base-graph $G^k$. \\ \hline 
$\phi :V\to G^k$ & Mapping function that achieves the mapping from $V$ onto $G^k$. \\ \hline 
$d\left(\phi \left(x\right),\phi \left(y\right)\right)$ & L1 distance between $\phi \left(x\right)$ and $\phi \left(y\right)$ in $G^k$. For   $\phi \left(x\right)=\left(j,k\right)$, $\phi \left(y\right)=\left(m,o\right)$ evaluated as\newline $d\left(\left(j,k\right),\left(m,o\right)\right)=\left|m-j\right|+\left|o-k\right|$. \\ \hline 
$p\left(\phi \left(x\right),\phi \left(y\right)\right)$ & The probability that $\phi \left(x\right)$ and $\phi \left(y\right)$ are connected through an $\text{L}_l$-level entanglement in $G^k$. \\ \hline 
$H_n$ & Normalizing term, defined as $H_n=\sum_z{d\left(\phi \left(x\right),\phi \left(z\right)\right)}$. \\ \hline 
$c_{\phi \left(x\right),\phi \left(y\right)}$ & Constant, defined as\newline $c_{\phi \left(x\right),\phi \left(y\right)}={\Pr }_{\text{L}_l}\left(E\left(x,y\right)\right)-\frac{d{\left(\phi \left(x\right),\phi \left(y\right)\right)}^{-k}}{H_n},$\newline where ${\Pr }_{\text{L}_l}\left(E\left(x,y\right)\right)$ is the probability that nodes $x,y\in V$ are connected through an $\text{L}_l$-level entanglement in the overlay quantum network $N$. \\ \hline 
$\Pr \left(\left.E\right|\phi \right)$ & Conditional probability between the $\phi \left(\cdot \right)$ configuration of positions of the quantum nodes in $G^k$ and the set $E$ of the $m$ edges of the overlay network $V$. \\ \hline 
$\Pr \left(\left.\phi \right|E\right)$ & Posteriori distribution of configuration $\phi $ at a given set $E$. \\ \hline 
$\Pr \left(\phi \right)$ & Candidate distribution. \\ \hline 
$q\left(\left.r\right|s\right)$ & Proposal density function to stabilize the Markov chain, proposes a next state $s^*$ given a state $s_i$. \\ \hline 
$u_j$ & The $j^{th}$ neighbor quantum node of $x_i$, $\left\{x_i,u_j\right\}\in E$ with base-graph position $\phi \left(u_j\right)\in G^k$. \\ \hline 
$v_j$ & The $j^{th}$ neighbor quantum node of $y_i$, $\left\{y_i,v_j\right\}\in E$ with base-graph position $\phi \left(v_j\right)\in G^k$ \\ \hline 
$\left|\left.\phi \left(u_j\right)\right\rangle \right.$, $\left|\left.\phi \left(v_j\right)\right\rangle \right.$ & Quantum systems, prepared locally by all $u_j$ and $v_j$ neighbor nodes of $x_i,y_i$. \\ \hline 
$M$ & Local measurement, which yields $M\left|\left.\phi \left(u_j\right)\right\rangle \right.=\phi \left(u_j\right)$ and $M\left|\left.\phi \left(v_j\right)\right\rangle \right.=\phi \left(v_j\right)$. \\ \hline 
$\zeta \left(x_i,y_i\right)$ & Parameter for the evaluation of the results of the local measurements of two nodes $x_i$ and $y_i$. \\ \hline 
$\Phi \left(x_i,y_i\right)$ & Parameter for the evaluation of the results of the local measurements of two nodes $x_i$ and $y_i$. \\ \hline 
swap & Swap operation. The $x_i,y_i$-swap of ${\phi }_1$, such that ${\phi }_1\left(x_i\right)={\phi }_2\left(y_i\right)$, ${\phi }_1\left(y_i\right)={\phi }_2\left(x_i\right)$, and ${\phi }_1\left(z_i\right)={\phi }_2\left(z_i\right)$ for all $z_i\ne x_i,y_i$. \\ \hline 
$p_{swap}\left(\phi \left(x_i\right),\phi \left(y_i\right)\right)$ & Swapping probability. Nodes $x_i,y_i$ swap their position information with this probability. \\ \hline 
$\mathcal{A}$ & Decentralized algorithm $\mathcal{A}$ in the $k$-dimensional $n$-sized base-graph $G^k$. \\ \hline 
$D\left(G^k\right)$ & Diameter of $G^k$. Refers to the maximum value of the shortest path (total number of edges on a path) between any pair of mapped nodes in $G^k$. \\ \hline 
$D\left(\mathcal{A}\right)$ & Minimal number of steps required by $\mathcal{A}$ to find the shortest path. \\ \hline 
$B_n$ & Box of size $n\times n$. \\ \hline 
$S_i$ & Subsquare of $B_n$ of side length $n^{\gamma }$, where ${k}/{4}<\gamma <1$. \\ \hline 
$S_{ik}$ & Sub-subsquares of side length $n^{{\gamma }^2}$, yielded from the subdivision of a subsquare $S_i$ into smaller units. \\ \hline 
$A_1$ & Event that there exists at least two subsquares $S_i$ and $S_j$ in $B_n$ such that there is no exists edge between them. \\ \hline 
$A_2$ & Event that there exists at one $S_i$ in $B_n$ such that there are two sub-subsquares $S_{ik}$ in $S_i$ which are not connected by edge. \\ \hline 
$D_{max}\left(S_i\right)$ & Largest diameter of the $S_i$ subsquares of side length $n^{\gamma }$. \\ \hline 
$D_{max}\left(S_{ik}\right)$ & The largest diameter of the $S_{ik}$ sub-subsquares of side length $n^{{\gamma }^2}$. \\ \hline 
$T\left({\phi }_1,{\phi }_2\right)$ & Transition matrix, where ${\phi }_2$ is the $x_i,y_i$-swap of ${\phi }_1$, such that ${\phi }_1\left(x_i\right)={\phi }_2\left(y_i\right)$, ${\phi }_1\left(y_i\right)={\phi }_2\left(x_i\right)$, and ${\phi }_1\left(z_i\right)={\phi }_2\left(z_i\right)$ for all $z_i\ne x_i,y_i$. \\ \hline 
$\Omega \left({\phi }_1,{\phi }_2\right)$ & Parameter for the definition of Markov chain.   \\ \hline 
$\varepsilon \left({\phi }_1,{\phi }_2\right)$ & Parameter for the definition of Markov chain. \\ \hline 
$E\left(x\vee y\right)$ & Edges connected to $x\in V$ or $y\in V$. \\ \hline 
$m$ & Iteration step. Utilizing the tessellation of $B_n$ for $m$ times results in end squares with side length $n^{{\gamma }^m}$, for which situation $m$ events, $A_1,\dots ,A_m$ exist. \\ \hline 
$C$, $Z$ & Constants, $C>0$, $Z>0$. \\ \hline 
$e_j$ & Event. \\ \hline 
$\Pr \left(e_j\right)$ & Probability that an event $e_j$ occurs. \\ \hline 
$X_j$ & Geometric random variable. \\ \hline 
$E\left(X_j\right)$ & Mean $E\left(X_j\right)$ of an geometric random variable $X_j$, evaluated as $E\left(X_j\right)=\frac{1}{\Pr \left(e_j\right)}=\mathcal{O}\left(\log n\right)$, where $n$ is the size of the $k$-dimensional base-graph $G^k$. \\ \hline 
\end{longtable}
\end{center}
\end{document}